\documentclass[aps,pra,twocolumn,superscriptaddress]{revtex4}
\usepackage{graphicx}
\usepackage{amssymb}
\usepackage{subfigure}
\usepackage{amsmath}
\usepackage{amsfonts}
\usepackage{bm}
\usepackage{color}
\newcommand{\beq}{\begin{equation}}
\newcommand{\eeq}{\end{equation}}
\newcommand{\vect}[1]{\ensuremath{\bm{{#1}}}}
\newcommand{\ket}[1]{\ensuremath{\left|{#1}\right\rangle}}
\newcommand{\bra}[1]{\ensuremath{\left\langle{#1}\right |}}

\newcommand{\oper}[1]{\mathbf{\mathsf{#1}}}

\newcommand{\sinc}{\ensuremath{\mathrm{sinc}}}

\newcommand{\spww}[1]{{\color{red}\small}}

\begin{document}


\title{Quantum Information Processing by Weaving Quantum Talbot Carpets}

\author{Osvaldo Jim\'enez Far{\'\i}as}
\affiliation{Centro Brasileiro de Pesquisas F\'{\i}sicas, Rio de Janeiro, RJ , Brazil}
\author{Fernando de Melo}
\affiliation{Centro Brasileiro de Pesquisas F\'{\i}sicas, Rio de Janeiro, RJ , Brazil}
\author{Perola Milman}
\affiliation{Laboratoire Mat\'eriaux et Ph\'enom\`enes Quantiques, Universit\'e Paris Diderot, CNRS UMR 7162, 75013, Paris, France}
\author{Stephen P. Walborn}
\email{swalborn@if.ufrj.br}
\affiliation{Instituto de F\'{\i}sica, Universidade Federal do Rio de
Janeiro, Caixa Postal 68528, Rio de Janeiro, RJ 21941-972, Brazil}

\begin{abstract}
Single photon interference due to passage through a periodic grating is considered in a novel proposal for processing $D$-dimensional quantum systems (quDits) encoded in the spatial degrees of freedom of light. We show that free space propagation naturally implements basic single quDit gates by means of the Talbot effect: an intricate time-space carpet of light in the near field diffraction regime.  Adding a diagonal phase gate, we show that a complete set of single quDit gates can be implemented. We then introduce a spatially-dependent beam splitter that allows implementation of controlled operations between two quDits. A new form of universal quantum information processing can then be implemented with linear optics and ancilla photons.  Though we consider photons, our scheme should be directly applicable to a number of other physical systems.  Interpretation of the Talbot effect as a quantum logic operation provides a beautiful and interesting way to visualize quantum computation through wave propagation and interference.     
\end{abstract}



\maketitle
\section{Introduction}

Quantum information (QI) science offers novel ways to transmit and process information, providing interesting advantages in terms of efficiency and security. As in the classical case, QI has focused on the processing of digital information instead of analog due advantages such as less demanding memory requirments and robustness against unavoidable errors. This has motivated an enormous quest for controllable quantum systems with discrete degrees of freedom in spaces of at most countable dimensions. 

In the history of QI, photonic systems have played a key role, offering several alternatives for encoding information.  The most common one is perhaps polarization of single photons, a true two-level quantum system. For $D>2$, a number of distinct features or advantages can appear, such as  tests of quantum contextuality \cite{cabello08}, quantum computation  \cite{gedik14}, higher-violation of Bell's inequality \cite{collins02,barbieri06} and increased information transmission rate and improved security in quantum key distribution \cite{bechmann00a,cerf02}. As such, additional photonic degrees of freedom, such as orbital angular momentum \cite{barreiro08,nagali09,dambrosio12}, transverse momentum \cite{neves05,walborn06a,tasca11}, time bins \cite{deRiedmatten04,ali-khan07}, either alone or in combination, have been used or proposed.  However, in general these higher-dimensional systems are typically more difficult to manipulate or control.  

Here we revisit two ``vintage'' concepts from optics and QI sciences to propose a new way of encoding and processing information in $D$-dimensional quantum systems. The first idea is to make use of the ``big space" available in a continuous (analog) degree of freedom in order to encode discrete quDits using periodic wave functions, as suggested by Gottesman-Kitaev-Preskill (GKP) \cite{gkp01}.  In our case these quDits can be prepared by sending light through a diffraction grating.  

The processing of these quDits will be realized in part using a concept dating back to 1836, when the Talbot effect was discovered by Henry Fox Talbot\cite{talbot1836}: in a standard diffraction experiment, light traversing a grating of period $l$ generates localized wave packets right after the grating. For rational propagation distances $ q/r$,  in units of twice the so-called Talbot distance $z_T=l^2/\lambda$, the initial state at the grating turns into a superposition of $r$ copies of the grating pattern, with each copy displaced transversely by a distance $l/r$. For irrational propagation distances, the diffraction pattern exhibits more complex fractal behavior \cite{berry96}. Mapping the intensity of light as a function of the transverse position and propagation distance displays an intricate pattern, known as a ``Talbot carpet of light". We show that, for proper encoding of quDits in the transverse spatial profile of single photons, certain instances of the Talbot effect are equivalent to quantum logical operations.  Placing phase elements at appropriate propagation distances is sufficient to implement arbitrary $D$-dimensional unitary operations. Historical reviews of the Talbot effect and its applications can be found in \cite{patorski89}. This phenomenon is not exclusive of optics, as it has also been observed in matter systems such as atomic beams \cite{chapman95} and as Bose Einstein condensates \cite{nowak97}.  As such, our results should be applicable to a number of physical systems.
 
 In section~\ref{sec:talbot} of this paper, we make a brief review of the Talbot effect. We then propose our encoding scheme for discrete orthonormal states in the continuous transverse distribution of single photons, reminiscent of the GKP encoding scheme \cite{gkp01}, and show how these states can be manipulated using the Talbot effect. This leads us to define \emph{Talbot quDits}. In section~\ref{sec:gates} we show that a complete set of single quDit gates can be realized by two optical primitives.  The first is near-field free-propagation that realizes the Talbot effect.  The second is an optical element that imprints a phase $\phi(x)$ on the spatial profile of the photons. When combined, these two operations lead to a natural implementation of the quantum Fourier transform (QFT).  In section \ref{sec:sdbs} propose a modified beam splitter that is capable of realizing two-photon gates when aided by post-selection. The combination of this two-quDits gate with the single quDit gates constitutes a universal set for quantum computation.    In sections \ref{sec:univ} and \ref{sec:error} we discuss the universality of this gate set and the resistance to errors.  We provide concluding remarks in section \ref{sec:conc}.

\section{Fractional Talbot Effect and Talbot qudits}
\label{sec:talbot}
 \begin{figure*}
  \includegraphics[width=\textwidth,height=7cm]{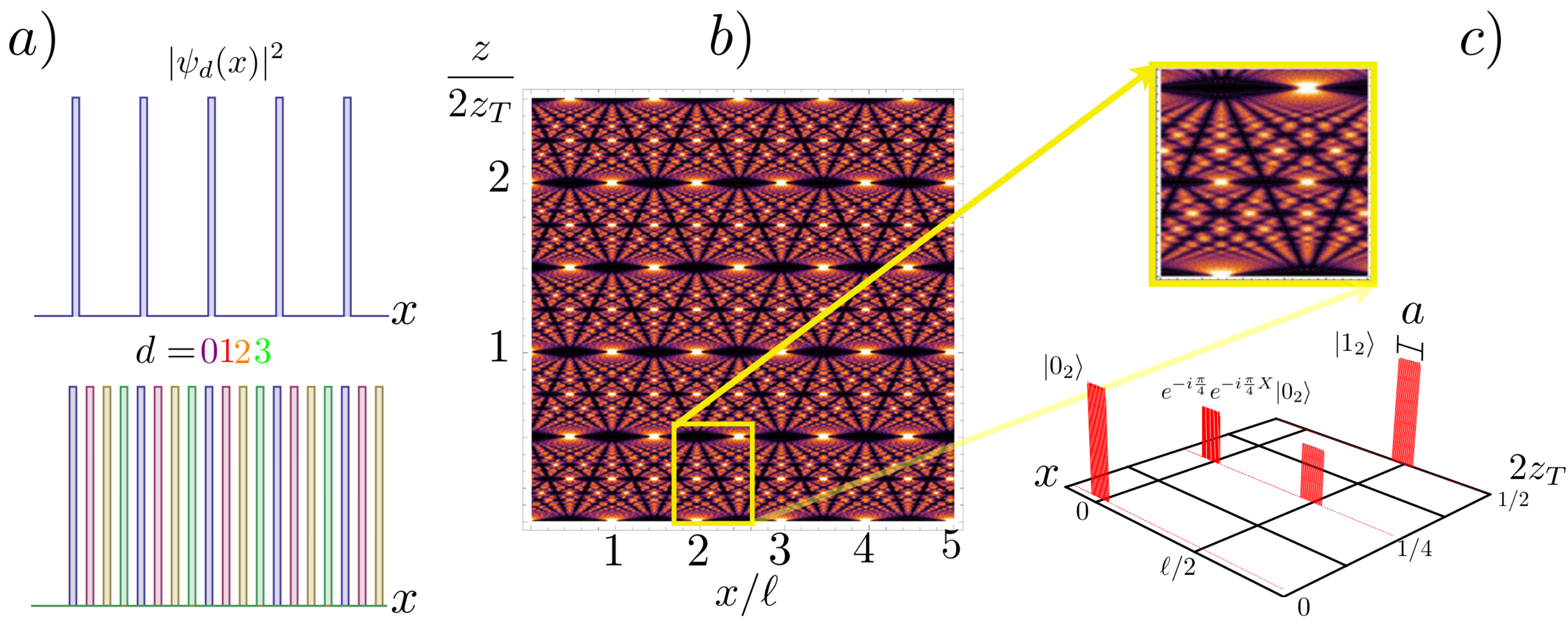}
  \caption{ In a) the intensity in the transverse direction of the periodic wave function which represents the encoding of a state $|d \rangle$  and an illustrations of the 4 different wave functions encoding the computational basis in dimension four, $\{0_4\rangle$, $|1_4\rangle$, $\{2_4\rangle$, $\{3_4\rangle\}$. In b) the diffraction pattern in the near field of a periodic grating in the z-x plane, the \emph{Talbot Carpet}; white regions indicate maximum intensity and black regions correspond to zero intensity. In c) a cell in the talbot carpet of of sides $\ell/2$ along $x$ and $z_T$ in $z$, where one can identify three states of a Talbot qubits along the propagation in $z$. \label{fig:encoding}}    
\end{figure*}

The Talbot effect  appears in the near field of diffraction of light from periodic gratings. It is characterized by the production of self-images and replicas of the grating for given propagation distances from the grating. Here we review some basic aspects of this phenomenon, following Ref. \cite{case09}.  Let us start by considering a plane wave with wavelength $\lambda$ propagating along the $z$-axis. At the plane $z=0$ is placed a grating with rectangular slits of width $a$ periodically separated by a distance $\ell$ along the $x$-axis. The slits are considered to be very large compared to $\lambda$ in the $y$ direction such that no diffraction occurs in this direction, and we can thus consider the effect occurring only in the $x$-$z$ plane. The wave at $z=0$ takes the form given by the transmission function of the grating, and can be written as a Fourier expansion    
\begin{equation}
\psi(x,0)=\psi_0(x)= \sum_m A_m e^{ i m xk_\ell}.
\label{eq:psix}
\end{equation}
That is, the grating imprints only multiples of $k_\ell=2\pi/\ell$ to the transverse momentum of the outgoing wave. The Fourier coefficients in this case are given by $A_m= (a/\ell)e^{-i \pi m a/\ell}\sinc(2m\pi a/\ell)$.  The top of Fig. \ref{fig:encoding} a) shows a pattern equivalent to $|\psi(x,0)|^2$.

Free space propagation simply imprints a phase $e^{izk_z}$ on the wave-function, with $z$ being the distance from the grating. An essential ingredient in the description of the Talbot effect is the paraxial approximation, for which the transverse wave vector has a very small contribution to the total wave vector $k=2\pi/\lambda=\sqrt{k_x^2+k_z^2}$ and then $k_z\approx k-\frac{(mk_\ell)^2}{2k}$ which leads to 
\begin{equation}
\psi(x,z)= \sum_m A_m e^{ i m xk_\ell}e^{izk_z} 
\approx \sum_mA_m e^{imxk_\ell}e^{-i\pi m^2z/z_T},
\label{eq:psitot}
\end{equation}
where the Talbot distance $z_T=\ell^2/\lambda$ has been defined, and we ignored a global phase. 

The recognition of the characteristic distance $z_T$ highlights the difference between the far field and near field diffraction. While in the far field diffraction the well known condition for constructive interference $\ell$sin$\theta=n\lambda$ dictates the position of the intensity peaks in the transverse plane (orders of diffraction), the near field constructive interference manifests itself in a totally different way.   
 Consider  propagation of the wave in equation (\ref{eq:psitot}) by distances 
 \begin{equation}
 z=(s+q/r)2z_T, 
 \label{z}
 \end{equation}
where $s$, $q$ and $r$ are integers, and $q$ and $r$ are relative primes. In this case, one can see that  
\begin{equation}
\psi(x,z) = \sum_{j=0}^{r-1} a_j \psi_0\left(x-\frac{j}{r} \ell \right), 
\label{eq:replic}
\end{equation}
where 
\begin{equation}
a_j = \frac{1}{r}\sum_{n=0}^{r-1} e^{-2 \pi i (n^2-jn)q/r}.
\label{eq:aj} 
\end{equation}
Equation (\ref{eq:replic}) shows that at distances spelled out by condition (\ref{z}), the wave-function is a linear combination of $r$ replicas of the initial state $\psi_0$ but with each copy displaced in the $x$ direction by integer multiples of $\ell/r$.  Coefficients $a_j$ define the weight and phase relation between the self-images, and their behaviour depending on $p$ and $r$ will be analysed later. As it can be noticed, equations \eqref{eq:replic} and \eqref{eq:aj} do not depend on $s$, which makes explicit the fact that the interference pattern is periodic in $z$ with period  $2z_T$. While the effect of self replication for $z=2z_T$ was first observed by Talbot in 1936 and explained theoretically by Raylegh in 1881, the whole interference pattern created by $|\psi(x,z)|^2$, known as the \emph{Talbot carpet of light}, to our knowledge was only observed in detail by Case \emph{et al.}~\cite{case09} in 2009. A computer simulation of a portion of a Talbot carpet is shown in Fig. \ref{fig:encoding} b). One can see that at integer values of $z/z_T$ the original wave function (z=0) is recovered, while at half-integer distances the original pattern is shifted by $\ell/2$. The Talbot effect has also being observed experimentally in matter waves with interfering atoms \cite{nowak97}, and Bose-Einstein condensates \cite{manfred11}. 

For distances not satisfying condition (\ref{z}), for example for irrational values of $z/2 z_T$ or $z=x$, the intensity pattern can display fractal behaviour as was extensively discussed in work developed in  \cite{berry96,berry296,hannay80}. Despite the richness of the physics involved in the weaving of the Talbot Carpet, in this work we restrict our analysis to the rational Talbot effect, that is, to those distances for which condition (\ref{z}) is satisfied.  

\medskip
\subsection{Talbot QuDits}
After this brief introduction to the Talbot effect, let us introduce ``Talbot quDits", which will be used to process $D$-dimensional quantum information.  In terms of the periodic wave function $\psi_0$, a $D$-dimensional quantum system can be encoded in the spatial distribution of single photons passing through a grating of period $\ell$. Each state $\ket{d}_D$, with $d=0,\ldots, D-1$, corresponds to a displacement by $\frac{d}{D}\ell$ of the initial comb $\psi_0(x)$.  The corresponding wave function is
\begin{equation}
\langle x\ket{d_D}=\psi_0\left(x-\frac{d}{D} \ell \right).
\label{eq:qudits}
\end{equation}

This encoding basically fills the space in the interval $\{0,\ell\}$ with wave functions centred at position $\frac{d}{D}\ell$ and then repeats periodically. To illustrate the idea, the bottom part of Fig. \ref{fig:encoding} b) shows the modulus square of the four basis states for $D=4$.  All the elements of the computational basis  are thus translations of the periodic comb $\psi_0$ by $\frac{m}{D}\ell$, with $m$ integer. That is, if we change variables  $x\rightarrow x+\frac{m}{D}\ell $ we obtain
\begin{equation}
\psi_0\left(x-\frac{d}{D} \ell \right)\rightarrow \psi_0\left(x-\frac{(d+m)}{D} \ell \right)=\langle x\ket{(d+m)_D},
\label{eq:qudits2}
\end{equation}
where the sum $(d+m)$ in $\langle x\ket{(d+m)_D}$ is modulo $D$. Notice that the state $|0_D\rangle$ is common to all dimensions and corresponds to the wave function $\psi_0(x)$ at the grating.

\begin{figure}
  \begin{center}
 \includegraphics[width=8cm]{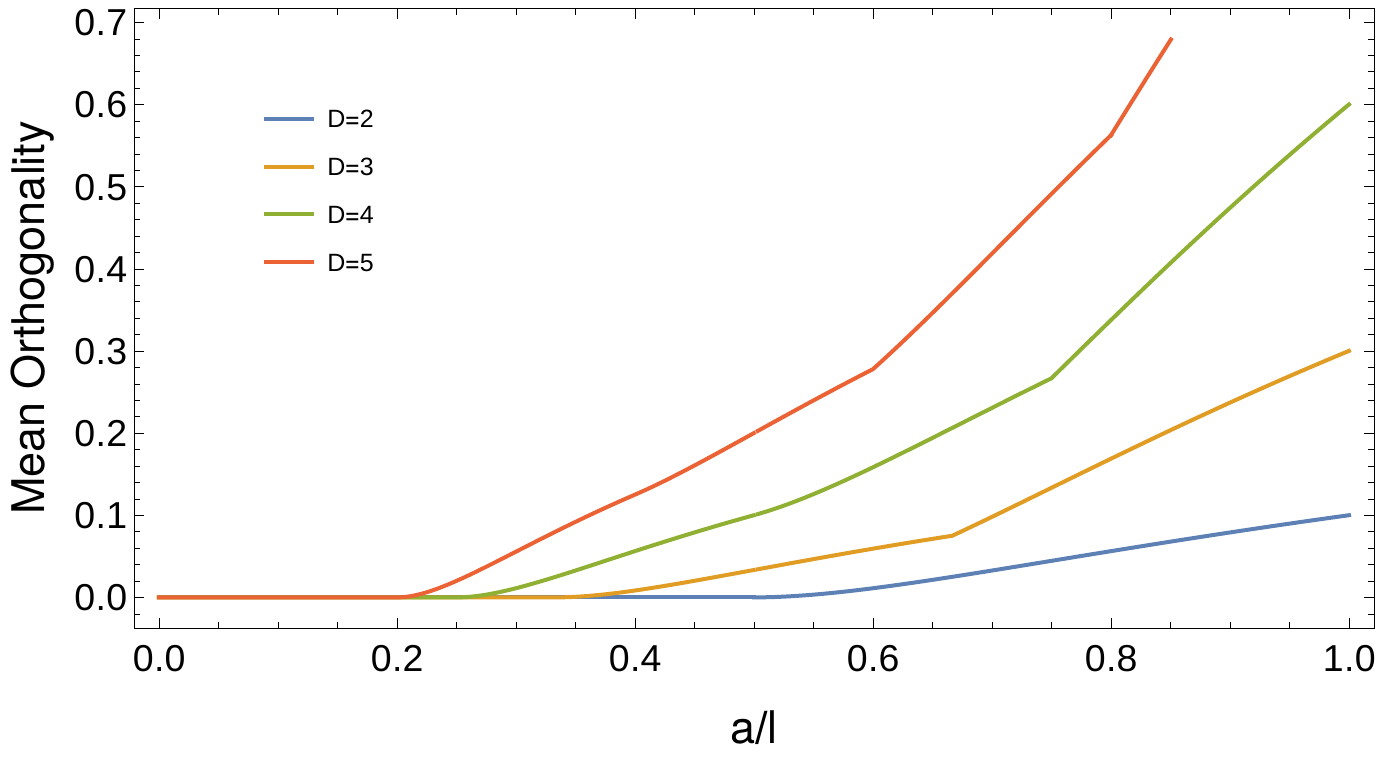}
  \caption{   
The plot shows the mean orthogonality $2 \sum_{d>d'}|\langle d_D\ket{{d'}_D}|^2/D(D-1)$ for the quDit computational states as function of $a/\ell$. For higher dimensions one needs the inter-slits distance $\ell$ to be much greater than the slit size $a$ in order to guarantee the orthogonality among the states. For fixed $a$ and $\ell$ the largest number of orthogonal states that one can encode is given by $D=\ell/a$, since in this way the space between two slits is evenly divided among non-overlapping states.}
\label{fig:orth}
\end{center}
  \end{figure}

 The orthogonality between two basis states, say $\ket{d_D}$ and $\ket{d'_D}$, will depend on the overlap $\langle d_D\ket{{d'}_D}$ between these wave packets which in turn depends on the ratio $a/\ell$. In Fig.~\ref{fig:orth} we show the mean orthogonality, defined as $2 \sum_{d>d'}|\langle d_D\ket{{d'}_D}|^2/D(D-1)$, between all the states of the computational basis for a given dimension $D$. For a $D$-dimensional system perfect orthogonality can be clearly achieved for $\frac{a}{\ell}\in\{0,\frac{1}{D}\}$.
\par
This QuDit encoding scheme is reminiscent of that of the GKP protocol for encoding a qubit ($D=2$) in a quantum harmonic oscillator \cite{gkp01} and also the encoding with modular variables proposed in \cite{vernazgris14}.  Using a QI perspective, we now show that the rational Talbot effect can be used to implement quantum logic operations on the Talbot quDits defined in (\ref{eq:qudits}). Near field propagation of the appropriate distance implements the unitary operator whose elements are the coefficients in \eqref{eq:aj}. In the next sections we describe a method for production and manipulation of optical Talbot qubits in order to achieve the necessary tools for universal quantum computation. 

\section{Talbot Gates, Diagonal Gates and the Quantum Fourier Transform}
\label{sec:gates}

In this section we show in detail the type of transformations that can be naturally implemented  on the physical quDits described in the last section. We will define two types of operations that are primitives sufficient to generate any unitary in the encoding space, and in particular the quantum Fourier transform. We will name these two primitive operations \emph{Talbot gates} and \emph{diagonal gates} respectively. Hereafter distances in the $z$ direction will be given in units of $2z_T$.

We start by formalizing the effect of free space propagation on the Talbot quDits defined in equation (\ref{eq:qudits}), thus defining the Talbot gate. The question we want to answer is: Given a fixed dimension $D$, what is the unitary transformation $\oper{U}_{q/r}$ associated with a propagation distance $z = q/r$?  That is, we are going to describe the unitary matrix satisfying
\begin{eqnarray}
\psi(x,q/r)=\langle x|\oper{U}_{q/r}|j_D\rangle. 
\end{eqnarray}

It will be helpful to study  the cases of even and odd dimension $D$ separately. 
\subsection{Talbot gate: $D$ even}
Let us first consider a propagation distance $z=1/r$, where $r$ is an even integer. According to Eq.(\ref{eq:replic}), the number of displaced replications is at most $r$ depending on whether all the coefficients $a_j's$ are different from zero or not. The expression determining the coefficients $a_j's$ is interesting and very well known, it is called a \emph{Gauss sum}. Gauss sums are relevant in number theory, and were applied for instance in the demonstration of the quadratic reciprocity law by Gauss himself \cite{ireland, gauss}. More recently, factorization protocols whose working mechanism is based in Gauss sums, were experimentally implemented \cite{wolk11, mehring07}. The natural appearance of Gauss sums in the the Talbot effect was observed by Hannay and Berry in \cite{hannay80}, in which reference they also provided an explicit evaluation of these sums. For us it is sufficient to know that for $r$ an even number, half of the $a_j$'s vanish.  It is convenient to define $r=2D$, and as such only $D$ Gauss sums are non zero. This establishes a relation between a given propagation distance and the dimension of the Hilbert space we are encoding. As here we are taking $D$ to be an even number, one can show~\cite{hannay80} that the non zero $a_j$'s are those for which $j$ is also an even number and we obtain for the non-vanishing coefficients

\begin{eqnarray}
a_{2d}=\frac{1}{\sqrt{D}}e^{-i\frac{\pi}{4}}e^{i\frac{\pi d^2}{D}}.\label{eq:as}
\end{eqnarray}
Then for $z=1/2D$ we can rewrite Eq.(\ref{eq:replic}) as
\begin{align}
\psi(x,z=1/2D)=&\langle x|\oper{U}_{1/2D}|0_D\rangle \nonumber \\
 =& \sum_{d=0}^{D-1} a_{2d} \psi_0\left(x-{d}{D} \ell \right), \label{eq:repares}
\end{align}
which in terms of state vectors can be written as:
\begin{eqnarray}
\oper{U}_{1/2D}\ket{0_D}=\sum_{d=0}^{D-1}a_{2d}\ket{d_D}. 
\end{eqnarray}
Now we evaluate the wave-function in Eq.~\eqref{eq:repares} at the point $x'=x-\frac{m}{D}\ell$:
\begin{align}
\psi(x-\frac{m}{D}\ell,z=1/2D)= \sum_{d=0}^{D-1} a_{2d} \psi_0\left(x-\frac{m+d}{D}\ell \right).
\end{align}
This is equivalent to the free evolution of a grating displaced by $m\ell/D$ for a distance $z=1/2D$, i.e., this gives the free evolution of the quDit state $\ket{m_D}$. In general one has:
\begin{eqnarray}
\oper{U}_{1/2D}\ket{m_D}=\sum_{d=0}^{D-1}a_{2d}\ket{(d+m)_D},\label{eq:umeio2}
\end{eqnarray}
where the sum $(d+m)$ is made modulo $D$.  That is, propagation over a distance $z=1/2D$ takes one of the computational basis states into an equally weighted superposition state, with coefficients given by Eq. \eqref{eq:as}. 

Using the generalized Pauli shift operator \cite{gkp01}, defined as $\oper{X}^{d}\ket{m_D}=\ket{(d+m)_D}$ with $\oper{X}^D=\oper{I}$ and $\oper{I}$ is the identity operator, we can write
\begin{eqnarray}
\oper{U}_{1/2D}\ket{m_D}=\sum_{d=0}^{D-1}a_{2d}\oper{X}^{d}\ket{m_D}.\label{eq:u2d}
\end{eqnarray}
Since the coefficients in equation (\ref{eq:as}) satisfy $\sum_{d=0}^{D-1}|a_{2d}|^2=1$, the unitarity of $\oper{U}_{1/2D}=\sum_{d=0}^{D-1}a_{2d}\oper{X}^d$ is guaranteed.

\subsection{Talbot gate: $D$ odd}
We can proceed as before to see that in the case of odd dimensions it is convenient to consider propagation distances $z=1/D$. In this case all the coefficients $a_j$ are non-zero. Indeed
\begin{eqnarray}
a_d=\frac{1}{\sqrt{D}}\left( \tfrac{2}{D} \right)e^{i\frac{\pi}{4}(D-1)}e^{i\frac{\pi(D+1)^2 d^2}{D}}
\end{eqnarray} 
where $\left( \tfrac{a}{b} \right)$ is the Jacobi symbol defined for odd integers $b$, and depending on the dimension it takes values $1$ or $-1$. Then the relevant operator for odd dimension is
\begin{eqnarray}
\oper{U}_{1/D}=\sum_{d=0}^{D-1}a_d\oper{X}^d.
\end{eqnarray} 
So, as for the even dimensional case, the Talbot gate is a polynomial in $\oper{X}$. The mathematical structure of the this operator and its relation with the discrete Fourier transform (DFT) will be explored in the next section.
\par
The Talbot gate thus takes the computational basis states $\ket{d_D}$ and transforms them into equally weighted superpositions of the basis states. It is important to note that while the appearance of the Talbot gate for even and odd dimension is similar, physically they are associated to distinct propagation distances $z=1/2D$ and $z=1/D$, respectively. This means that the propagation distances for manipulation of the Talbot quDits depends on the parity of the chosen dimension.

\subsection{General properties of $\oper{U}_{1/r}$}
It is interesting to notice that the operator $\oper{U}_{1/r}$ is determined by the set of $D$  numbers $\vect{a}=\{a_0, a_2, ..., a_{(D-1)} \}$. The rows of the matrix associated to  $\oper{U}_{1/2D}$ are successive cyclic permutations of this vector: 
\begin{eqnarray}
\oper{U}_{1/r}=\left(
\begin{array}{ccccc}
a_0 & a_{D-1} & .& . & a_1\\
a_1 & a_0 & . & . & a_2\\
. & . & . & . & .\\
a_{D-1} & a_{D-2}&.&.&a_0\\
\end{array}
\right).
\end{eqnarray}
These types of matrices are known as \emph{circulant matrices} \cite{davis94}, and are closely related to the quantum Fourier transform (QFT) defined by
\begin{equation}
\oper{F}\ket{j_D}=\sum_{d=0}^{D-1}e^{-2 i\pi\frac{j d}{D}}\ket{d_D},
\label{eq:F}
\end{equation} 
since the latter diagonalizes any circulant matrix. Later on in this section we will discover a closer connection between $\oper{U}_{1/2D}$ and the QFT.
We can immediately observe some straightforward properties of the Talbot gates namely
\begin{eqnarray}
[(\oper{U}_{1/r})^a,(\oper{U}_{1/r})^b]=0
\end{eqnarray}
for $a$ and $b$ arbitrary integers. Also we can now construct the unitary matrix associated to a distance $z=q/r$ as powers of $\oper{U}_{1/r}$, that is $\oper{U}_{q/r}=\oper{U}_{1/r}^q$. In general, we can see that $\oper{U}_{a/r}\oper{U}_{b/r}=\oper{U}_{(a+b)/r}$ where all of the previous matrices are circulant matrices, as the product of circulant matrices is another circulant matrix. This allows to verify consistancy with the integer Talbot effect, which demands that 
\begin{eqnarray}
(\oper{U}_{1/r})^{r}=\oper{I}.
\end{eqnarray}
Moreover, when $D$ is even it is possible to check that $\oper{U}_{1/2D}^D=\oper{X}^{D/2}$, which shows the transverse shift of the intensity pattern at a propagation distance $z=1/2$, as observed in the Talbot effect, and illustrated in Fig. \ref{fig:encoding} b).

\subsection{Diagonal gates and the quantum Fourier transform}
\begin{figure}
  \includegraphics[width=8cm]{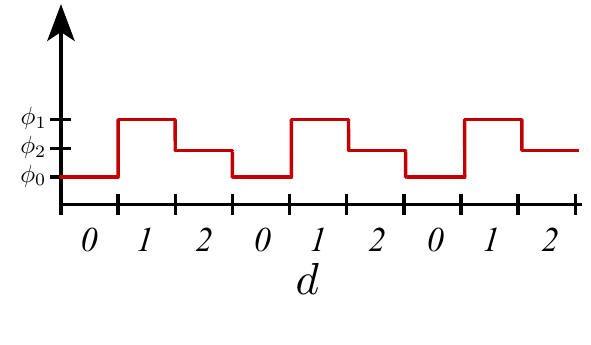}
  \caption{Example of the phase function $\vec{\phi_d}=(\phi_0,\phi_1,\phi_3$) for $D=3$.} \label{fig:phasegate}
\end{figure}
While free space propagation naturally implements   a whole family of single quDit transformations via the Talbot effect, some commonly used linear optics elements can be exploited to implement additional sets of important gates. We appeal to a spatial light modulator, or equivalent device, to imprint an $x$-dependent phase on the optical field.  As such it is straightforward to implement
\begin{equation}
\Psi(x) \longrightarrow e^{i \phi(x)} \Psi(x).
\end{equation}
Moreover, we can define the phase $\phi(x)$ to have distinct values $\phi_d$ in the regions $R_d(x)$ where $\psi_d(x)$ is non-zero.  An example of the phase function for $D=3$ is shown in Fig. \ref{fig:phasegate}.  Thus, we can perform
\begin{equation}
\ket{d_D} \longrightarrow e^{i \phi_d} \ket{d_D}.
\end{equation}
More generally, for a superposition state $\ket{\Psi} =  \sum_{d=0}^{D-1} \lambda_d  \ket{d_D}$,
\begin{equation}
\ket{\Psi} \longrightarrow  \sum_{d=0}^{D-1} \lambda_d e^{i \phi_d} \ket{d_D}.
\end{equation}
This is a quantum gate that is diagonal in the $\ket{d_D}$ basis, and can be written as
\begin{equation}
\oper{Z}_{\vec{\phi}} =  \sum_{d=0}^{D-1}  e^{i \phi_d} \ket{d_D}\bra{d_D}, \label{eq:diagonal}
\end{equation}
where $\vec{\phi} = (\phi_0,\dots,\phi_{D-1})$ represents the possible phase values. For $D=2$ the $\oper{Z}_{\vec{\phi}}$ represents a rotation by $\phi$ along the the $z$ axis in the Bloch's sphere. 
\par
The tools developed until now are useful to expose the relation between $Z_{\vec{\phi}}$ and $U_{1/r}$ to the QFT. Consider first the case of even dimension  $D$ and the diagonal gate setting $\vec{\phi}$ to $\vec{\Xi}$, for which the $d^{th}$ element reads $[\vec{\Xi}]_d=e^{i(\frac{\pi}{8}-\frac{\pi d^2}{D})}$. That is
\begin{eqnarray}
\oper{Z}_{\vec{\Xi}}=e^{\frac{i\pi}{8}}\sum_{d=0}^{D-1} e^{-\frac{i\pi d^2}{D}}\ket{d_D}\bra{d_D},
\end{eqnarray}
imprints a quadratic phase to each element of the basis. After some algebraic manipulation one obtains the matrix identity 
\begin{equation}
\oper{Z}_{\vec{\Xi}}\oper{U}_{1/2D} \oper{Z}_{\vec{\Xi}}\ket{j_D}=\sum_{d=0}^{D-1}e^{2\pi\frac{(j+d)j}{D}}\ket{(j+d)_D}=\oper{F}\ket{j_D},
\label{eq:Fpara}
\end{equation}
where $\oper{F}$ is the QFT operator defined in Eq. \eqref{eq:F}. That is, the operators $\oper{Z}_{\vec{\Xi}}$ and $\oper{U}_{1/2D}$ provide a decomposition of the QFT.  

Another interesting identity can be obtained by exploiting some known results for circulant matrices\cite{davis94}: First that diagonal gates $\oper{Z}_{\vec{\phi}}$ when conjugated by a Fourier transformation give a circulant matrix, i.e., $\oper{F} \oper{Z}_{\vec{\phi}} \oper{F}^{\dagger}=\oper{C}$ with $\oper{C}$ some circulant matrix. Second the defining property of circulant matrices which says that circulant matrices are diagonalized by Fourier transformations. Mathematically $\oper{F}\oper{U}_{1/r}\oper{F}^{\dagger}=\oper{Z}_{\vec{\gamma}}$ for some $\vec{\gamma}$.  Putting these two properties together with unitarity, $\oper{F}\oper{F}^{\dagger}=\oper{I}$, we can obtain another decomposition of the QFT:
\begin{eqnarray}
\oper{C}\oper{Z}_{\vec{\gamma}}\oper{C}=\oper{F}, \label{eq:Fparb}
\end{eqnarray}
This form of the QFT will appear later in the example of $D=2$.

For odd dimension we obtain the following decomposition:
\begin{eqnarray}
\oper{Z}_{\vec{\Theta}_{-}}\oper{U}_{D}\oper{Z}_{\vec{\Theta}_{+}}=\oper{F} \label{eq:Fodda}
\end{eqnarray}
where 
\begin{eqnarray}
\oper{Z}_{\vec{\Theta}_{-}}=\sqrt{\left(\frac{2}{D}\right)}\sum_{d=0}^{D-1}e^{-\frac{i\pi (d^2-D d)}{D}}|d_D\rangle\langle d_D|, \nonumber\\
\oper{Z}_{\vec{\Theta}_{+}}=\sqrt{\left(\frac{2}{D}\right)}\sum_{d=0}^{D-1}e^{-\frac{i\pi (d^2+D d)}{D}}|d_D\rangle\langle d_D|.
\end{eqnarray}
\subsection{Performing arbitrary unitary operations on Talbot QuDits}
\begin{figure}
  \includegraphics[scale=0.55]{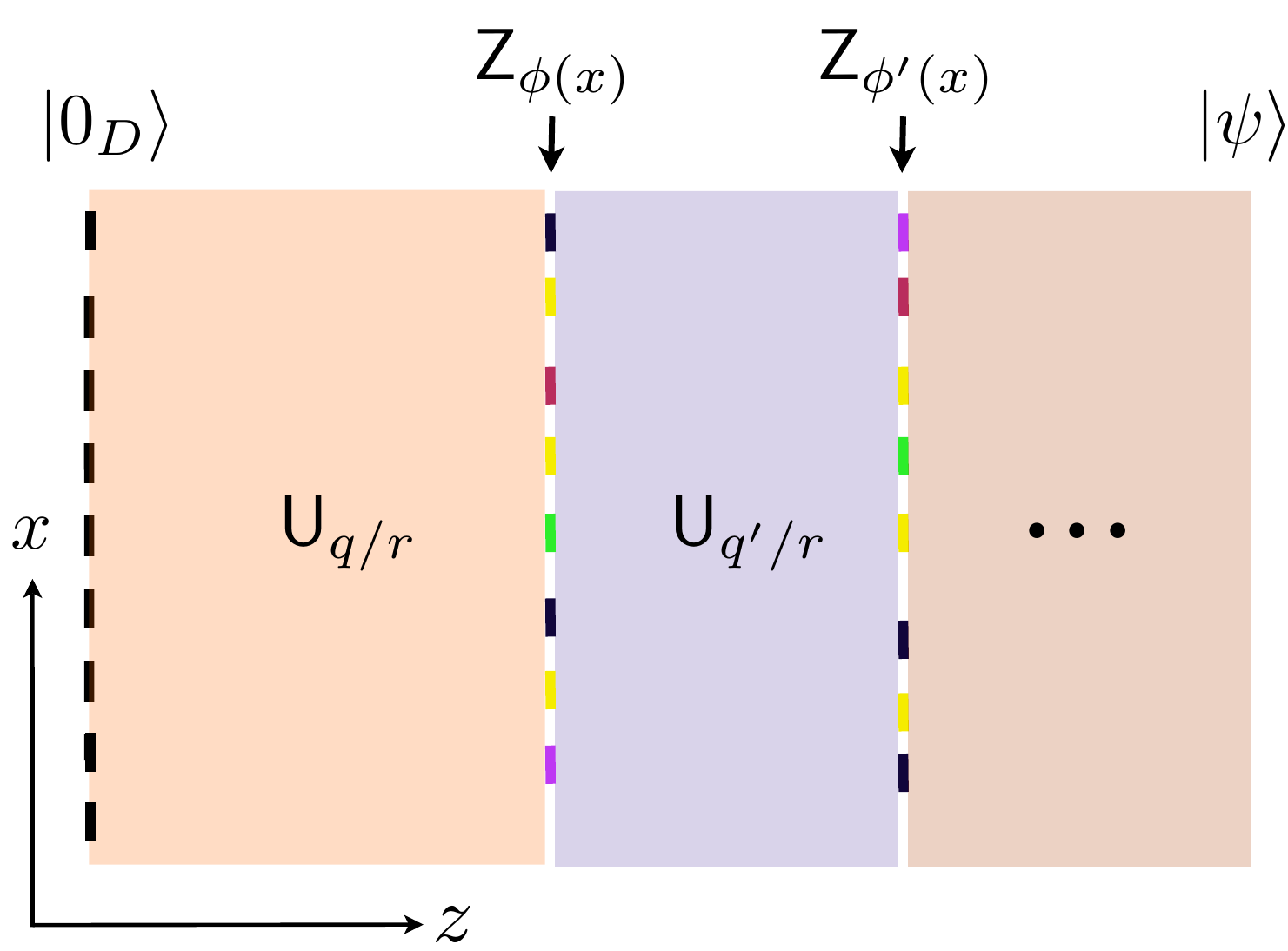}
  \caption{ The general scheme for processing quantum information with a Talbot quDit. Staring with a $|0_D\rangle$ at the grating, free space propagation by distance $q/r$ implements $\oper{U}_{q/p}$, until a diagonal gate is applied at $q/r$. Repeated applications of this sequence generate arbitrary unitary transformations in $SU(D)$ for $D$ prime}.\label{fig3}
\end{figure}
As illustrated in Fig.~\ref{fig3}, information processing with Talbot quDits is done by interlacing Talbot gates --- free propagations $\oper{U}_{q/r}$ --- with diagonal gates --- $\oper{Z}_{\vec{\phi}}$ with appropriate phase vectors $\vec{\phi}$ acting at positions $z=q/r$. It is important to notice that this set of operations lead to an immediate implementation of the quDit Clifford group, which contains $\oper{Z}_{\vec{\Omega}}$, with $[\vec{\Omega}]_d =e^{i \pi d(d-1)/D} $ and the $D$-dimensional discrete Fourier transform~\cite{gottesman1999fault}. This set of gates is very important for fault-tolerant implementations of quantum computation. Moreover, as shown in~\cite{campbell12}, if one adds to the Clifford gates a single arbitrary non-Clifford gate, which in our case can be any other diagonal gate, it is possible to approximate any unitary in $SU(D)$ to any desirable precision for $D$ odd prime. For $D=2$ we get an exact cover of $SU(2)$, as we now show. 

\subsection{Example: Talbot qubits}

In order to illustrate the results of last section we take the important case of Talbot qubits, i.e., $D=2$. According to Eq.(\ref{eq:u2d}), we can write the unitary operator associated to a propagation $z=1/4$ as
\begin{eqnarray}
\oper{U}_{1/4}=\frac{e^{-i\frac{\pi}{4}}}{\sqrt{2}}(\oper{I}+i \oper{X}),
\end{eqnarray} 
where the Pauli operator satisfies $\oper{X}^2=\oper{I}$. Using the identity $e^{i\theta \oper{X}}= \cos(\theta)\oper{I}+i\sin(\theta)\oper{X}$, one notices some useful relations 
\begin{align}
&\oper{U}_{1/4}=e^{-i\frac{\pi}{4}}e^{i\frac{\pi}{4}\oper{X}} \\
&\oper{U}_{1/2}=(\oper{U}_{1/4})^2=\oper{X}\\
&\oper{U}_{1}=(\oper{U}_{1/2})^2=\oper{I}.
\end{align}
It is then clear that $\oper{U}_{1/4}$ takes the form of a rotation in the Bloch sphere by $\frac{\pi}{4}$ along the $x$ axis times a constant phase of $-\pi/4$.

\begin{figure}
  \begin{center}
 \includegraphics[scale=0.43]{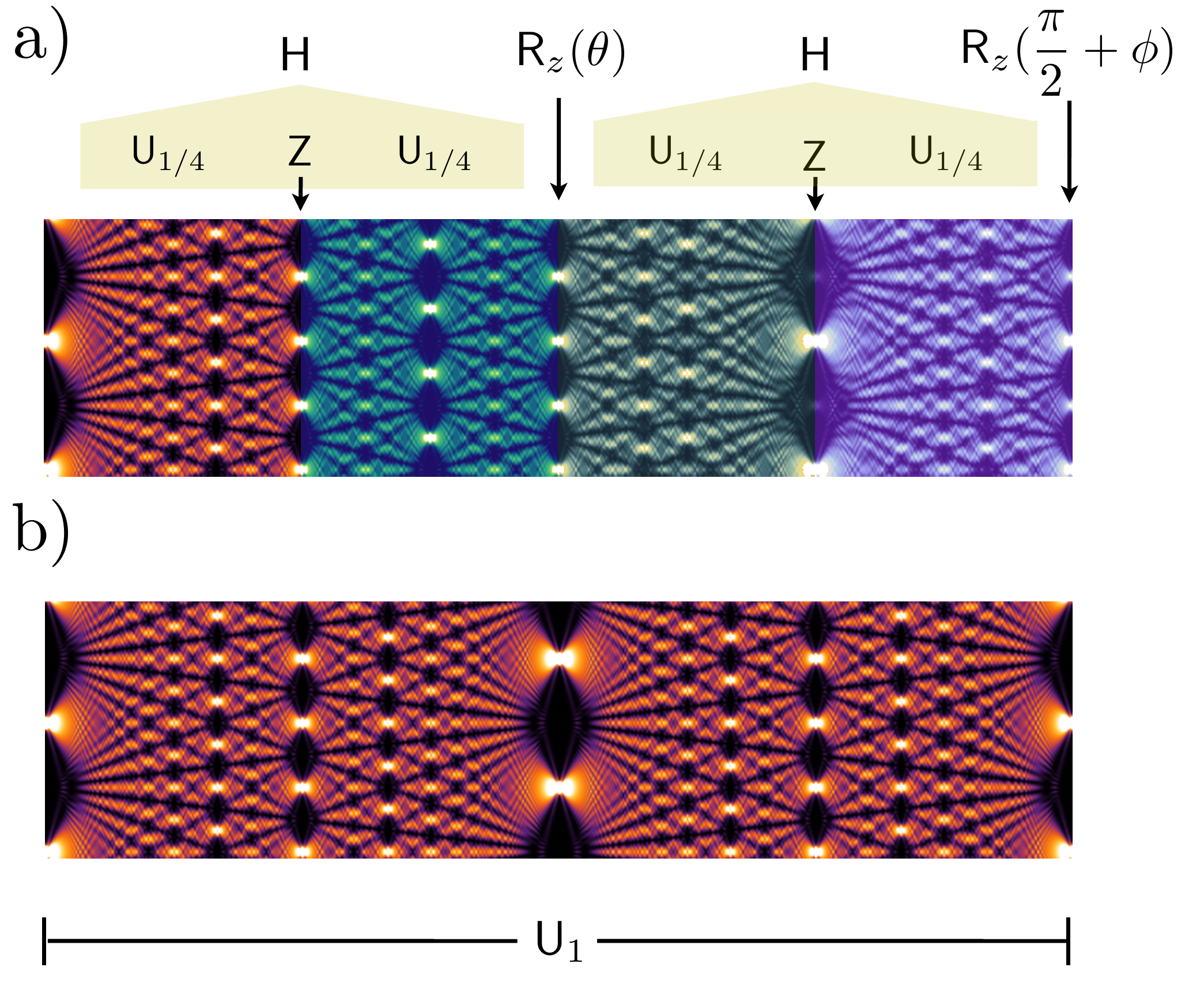}
  \caption{  In a) we show the circuit consisting Hadamard gates and rotations in the Bloch sphere along the z axis needed to prepare arbitrary qubit states. In this case we show the preparation of a state defined by the polar and azimuthal angles $(\theta, \phi)=(\pi/2.7, \pi/2)$  in the Bloch sphere. The color changes are to illustrate that a phase gate has been applied at the regions marked with vertical arrows. This circuit requires $2 z_T$ of propagation in order to complete the state preparation. For comparison we show the Talbot carpet due to free evolution throughout the same distance $ 2z_T$, which thus implements the identity operator.}
\label{fig:statePrep}
\end{center}
  \end{figure}

With the Talbot gates $\oper{U}_{1/4}$, $\oper{U}_{1/2}$ and the set of gates $\oper{Z}_{\vec{\phi}}$, it is possible to construct other interesting operators. Consider for instance the operator identity $e^{-i \frac{\pi}{2}}e^{i\frac{\pi}{4}\oper{X}}e^{i\frac{\pi}{4}\oper{Z}}e^{i\frac{\pi}{4}\oper{X}}=\oper{H}$ where $\oper{H}$ is the Hadamard gate. It is easy to see that
\begin{equation}
\oper{H}=\oper{U}_{1/4}\oper{Z}_{\pi/4}\oper{U}_{1/4}, \label{eq:hada}
\end{equation} 
where $\oper{Z}_{\pi/4}= e^{i \pi/4}\ket{0_2}\!\bra{0_2}+e^{-i \pi/4}\ket{1_2}\!\bra{1_2}$.
This last equation shows how it is possible to construct a Hadamard gate in a Talbot length with a single intervention of the diagonal phase  gate \eqref{eq:diagonal}. In the first half of Fig.~\ref{fig:statePrep} a) we illustrate the Talbot carpet associated to the realization of the Hadamard gate using Eq.\eqref{eq:hada}. This decomposition of the Hadamard gate is of the form of Eq.~\eqref{eq:Fparb}, and thus it is also possible to implement the same $\oper{H}$ gate in the form $\oper{Z}_{\vec{\Xi}}\oper{U}_{1/4}\oper{Z}_{\vec{\Xi}}$, which implements $\oper{H}$ in half a Talbot distance. 
 The Hadamard gate is very useful since it helps to prepare arbitrary unitaries, and as such to prepare any pure state. This is obtained  in conjunction with rotations along the $z$ axis by means of the identity \cite{chuang00}:
\begin{eqnarray}
\oper{R}_{z}\left(\frac{\pi}{2}+\phi\right)\oper{H}\oper{R}_z(\theta)\oper{H}|0\rangle=\rm{cos}(\theta)|0\rangle+e^{i\phi}\rm{sin}(\theta)|1\rangle.
\end{eqnarray}
Here the rotation operators $\oper{R}_{z}(\theta)$ are simply the diagonal phase gates $\oper{Z}_{\theta}$, given by
\begin{equation}
\oper{Z}_{\theta}= e^{i \theta}\ket{0_2}\!\bra{0_2}+e^{-i \theta}\ket{1_2}\!\bra{1_2}. 
\end{equation}

Fig \ref{fig:statePrep} a) shows an example of  the Talbot carpet produced in preparation of an arbitrary pure state from the initial state $\ket{0_2}$ for polar and azimuthal angles $(\theta, \phi)=(\pi/2.7, \pi/2)$  in the Bloch sphere.  As a comparision,  Fig \ref{fig:statePrep} b) shows the Talbot carpet produced for evolution under the operator $\oper{U}_1$, corresponding to the identity operator.

\section{Measurement}
Measurement of the Talbot QuDits in the computational basis can be performed simply by using an array of single photon detectors, single-photon sensitive cameras \cite{edgar12,aspden13}, or spatial light modulators \cite{paul14}.  In all of these techniques, registration of a photon at some spatial position indicates one of the basis states $\ket{d_D}$.  In the next section, we propose another method for measurement of computational states using a spatial analog of a polarizing beam splitter.      

\section{Spatially Dependent Beam Splitter}
\label{sec:sdbs}
\begin{figure}
  \begin{center}
 \includegraphics[width=7cm]{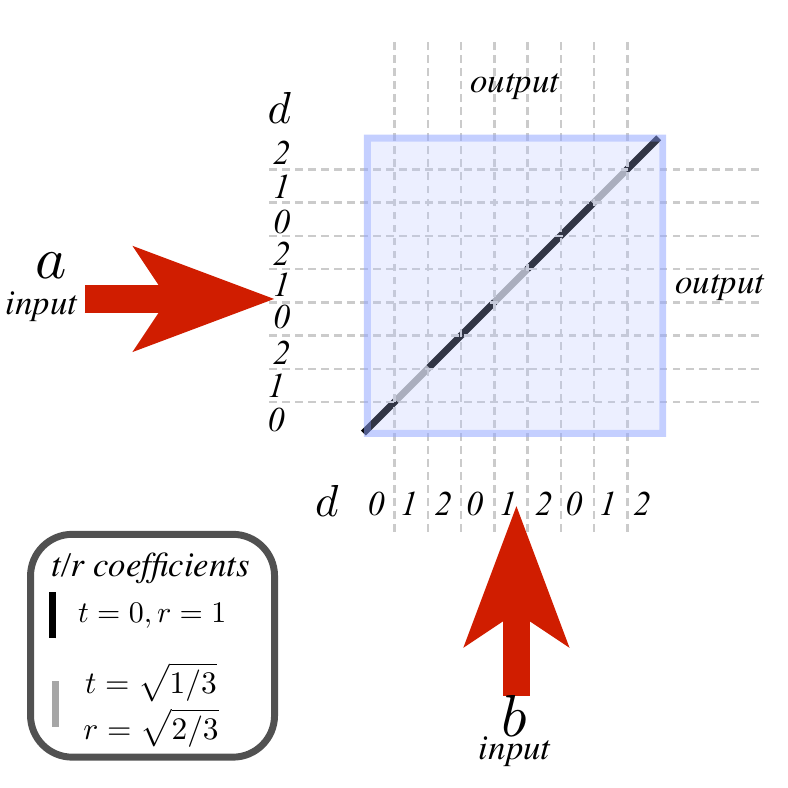}
  \caption{Example of a spatially dependent beam splitter (SDBS) for spatial quDits.  Shown is an example for $D=3$.  The colored regions represent parts of the beam splitter with different transmission coefficients $t_d$ and reflection coefficients $r_d$.}
\label{fig:SDBS}
\end{center}
  \end{figure}
We now propose a novel optical beam splitter  that will be useful for correlating different Talbot quDits.  The usual symmetric beam splitter (BS) has reflection and transmission coefficients $t$ and $r$, which in general are complex numbers such that $|t|^2+|r|^2=1$.  Consider now a spatially-dependent beam splitter, which we denote SDBS, such that the transmission and reflection coefficients depend on the spatial coordinate $x$.  For simplicity, we consider a single spatial dimension, however, the same device could be conceived for two spatial dimensions.  For each position $x$ we have transmission and reflection coefficients $t(x)$ and $r(x)$, respectively, such that $|t(x)|^2+|r(x)|^2=1\,  \forall \,  x$.        

To apply the SDBS device to our spatial quDits, we consider that we have distinct coefficients $t_d$ and $r_d$ corresponding to each periodic region $R_d(x)$ corresponding to the non-null regions of the computational basis wave functions $\psi_d(x)$, as illustrated in Figure \ref{fig:encoding}.   Illustration of such a device is shown in Fig. \ref{fig:SDBS}, for $D=3$. Fabrication of such a device should be feasible with evaporative coating techniques which are typically used in fabrication of mirrors, beam splitters and other optical devices. 

Consider as a first example the case in which $t_k=0$, $r_k=1$ for some specific $d=k$ and $t_d=1$, $r_d=0$ for all other values of $d \in\{0,\ldots, D-1\}$.  In this case, the SDBS acts as a filter for the computational basis states, as it reflects the computational basis state $\ket{k}$ and transmits all others $\ket{d}$.  This is analogous to the usual polarizing beam splitter, which typically reflects vertical polarization and transmits the horizontal polarization state.  Using a series of these devices, one can filter out all the basis states, and perform a projective measurement.

For the more general case, the SDBS transforms the Talbot QuDit basis states as
\begin{subequations}
\begin{align}
 |d_D\rangle_a & \longrightarrow t_d |d_D\rangle_a + i r_d |d_D\rangle_b \\
|d_D\rangle_b & \longrightarrow t_d |d_D\rangle_b + i r_d |d_D\rangle_a. 
\end{align}
\end{subequations}     
It is interesting to consider two-photon interference \cite{hom87}, where one photon is input in mode $a$ and the other in mode $b$.  If the photons are indistinguishable in all other degrees of freedom, and are prepared in the same computational basis state, we have 
\beq
|d_D\rangle_a |d_D\rangle_b \longrightarrow (t_d^2-r_d^2) |d_D\rangle_a |d_D\rangle_b + i t_d r_d (|d_D\rangle_a^2 +  |d_D\rangle_b^2). 
\eeq
If the photons are prepared in different computational basis states $\ket{d}$ and $\ket{f}$, but otherwise indistinguishable, we have 
\begin{align}
|d_D\rangle_a |f_D\rangle_b \longrightarrow & t_d^2|d_D\rangle_a |f_D\rangle_b -r_d^2 |f_D\rangle_a |d_D\rangle_b \nonumber \\
& + i t_d r_d (|d_D\rangle_a|f_D\rangle_a +  |d_D\rangle_b |f_D\rangle_b).  
\end{align}
\subsection{Two-quDit Gates}
\begin{figure}
  \begin{center}
 \includegraphics[width=7cm]{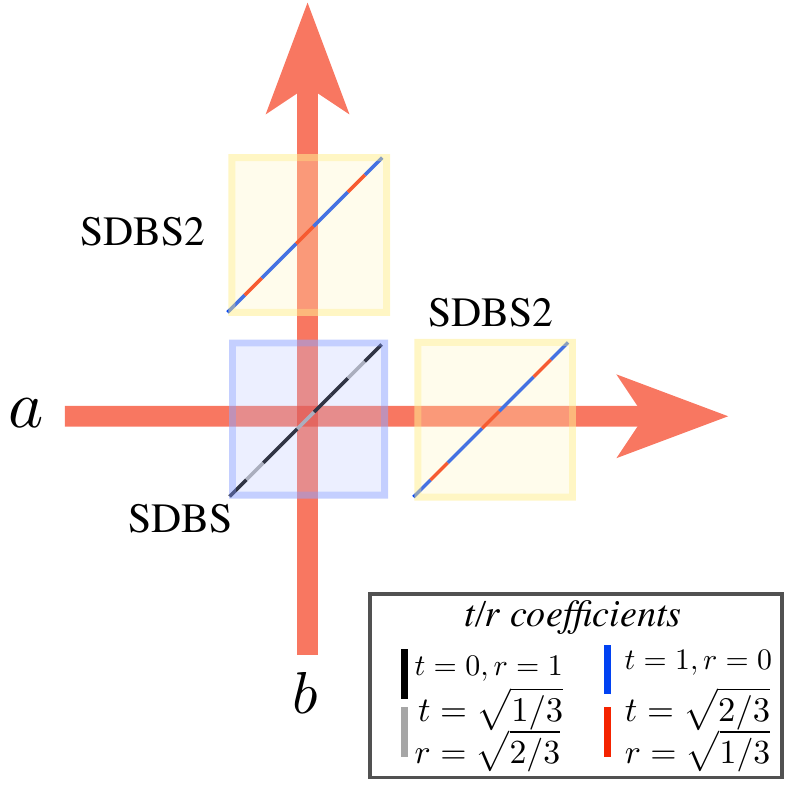}
  \caption{Several SDBS devices can be used to implement a controlled gate between Talbot QuDits (see text). SDBS and SDBS are spatially dependent beam splitters with different transmission and reflection coefficients.}
\label{fig:CPhase}
\end{center}
  \end{figure}
Two-photon gates based on linear optics were first proposed in the seminal paper by Knill, Laflamme and Milburn \cite{klm01}.  Additional two-photon gates were later proposed \cite{ralph01,ralph02}.   In particular, we consider first the scheme proposed in Ref.\cite{ralph02}, improved upon in Ref.\cite{hofmann02} and experimentally demonstrated in Refs.\cite{langford05,kiesel05,okamoto05}.  We will show that we can perform a two-photon phase gate for spatial Talbot quDits via post-selection, as shown in Figure~\ref{fig:CPhase}.  To do so, we consider two-photon interference on a first SDBS device, and define $t_k=1/\sqrt{3}$, $r_k=\sqrt{2/3}$ for a fixed basis state $k$, and $t_d=0$, $r_d=1$ for all other values of $d \in\{0,\ldots,k-1,k+1, \ldots,D-1\}$. Additional SDBS devices with coefficients $t_k=1$, $r_k=0$ for fixed basis state $k$, and $t_d=1/\sqrt{3}$, $r_d=\sqrt{2/3}$ for $d \in\{0,\ldots,k-1,k+1, \ldots,D-1\}$ are then used in modes $a$ and $b$ simply as filters. We will consider only events where one photon leaves each output of the final SDBS in each mode.  Then, all different combinations of computational basis states result in one of the following:
 \begin{subequations}
 \begin{align}
|k_D\rangle_a |k_D\rangle_b & \longrightarrow -\frac{1}{3} |k_D\rangle_a |k_D\rangle_b & \longrightarrow -\frac{1}{3} |k_D\rangle_a |k_D\rangle_b, \\ 
|k_D\rangle_a |d_D\rangle_b & \longrightarrow \frac{1}{\sqrt{3}} |k_D\rangle_a |d_D\rangle_b & \longrightarrow \frac{1}{3} |k_D\rangle_a |d_D\rangle_b, \\ 
|d_D\rangle_a |k_D\rangle_b & \longrightarrow \frac{1}{\sqrt{3}} |d_D\rangle_a |k_D\rangle_b & \longrightarrow \frac{1}{3} |d_D\rangle_a |k_D\rangle_b, \\ 
|d_D\rangle_a |d_D\rangle_b & \longrightarrow  |d_D\rangle_a |d_D\rangle_b & \longrightarrow  \frac{1}{3} |d_D\rangle_a |d_D\rangle_b, 
\end{align}
\end{subequations}
where the first column is the input state, the second column the state after two-photon interference at the first SDBS (and post-selection), and the third column the state after the filter SDBS (and post-selection).  We can see that the result is a $\pi$ phase shift when the two photons are in state $k$, and no phase shift for all other possibilities.  This is an entangling gate.  For $D=2$, this is equivalent to a controlled-phase gate, $\oper{CZ}_{\vec{0}}$, which is capable of creating a maximally entangled state \cite{ralph02,hofmann02}.    

Independent of the dimension $D$, we see that this gate operates with a success probability of $1/9$, due to the post-selection.  To make this gate scalable, post-selection with a non-demolition measurement of photon-number is required to ensure that one photon leaves each mode, as is the case for the gate based on photon polarization \cite{ralph02}.   


\section{Universality of Talbot carpet weaving}
\label{sec:univ}

For a computation model to be called \emph{universal} it must be the case that, for any fixed number $N$ of quDits, any unitary operation in $SU(D^N)$  can be decomposed, either exactly or up to any desirable precision, in terms of the gates that can be implemented within the model. If a set of gates is sufficient for universality it is called a universal gate set. In the previous sections we showed how to implement single- and two-quDit gates for Talbot quDits. Now we argue that these operations are sufficient for implementing universal quantum computation for prime dimensions.

For Talbot qubits, i.e. $D=2$, we showed how any single-qubit unitary and two-qubit controlled phase gate can  be implemented. The set of gates $\{\oper{Z}_{\vec{\phi}},\oper{H},\oper{CZ}_{\vec{0}} | \; \forall \; [\vec{\phi}]_0,[\vec{\phi}]_1 \in [0,2\pi] \}$ is a well known set for exact universal quantum computation with qubits~\cite{barrenco95}.

For Talbot quDits, with $D$ an odd prime, we have already seen that the implementation of discrete Fourier transformation and arbitrary phase gates allows for, at least, a dense cover of $D$-dimensional unitaries. The proposed spatially dependent beam splitter of the previous section give us an entangling gate, which, according to the results in Ref.~\cite{quDitsUni}, is sufficient to compose a universal set of gates. 

Recent results on decomposition of arbitrary unitary matrices in $SU(D)$ as the product of phase gates and discrete Fourier transformations acting on subspaces of varied dimension~\cite{idel14} might render our quantum computation model universal for all dimensions. However, the possibility of implementing a QFT gate in a subspace of a quDit still has to be further analysed.  We leave this for future work.

\begin{figure}[t!]
\includegraphics[width=\linewidth]{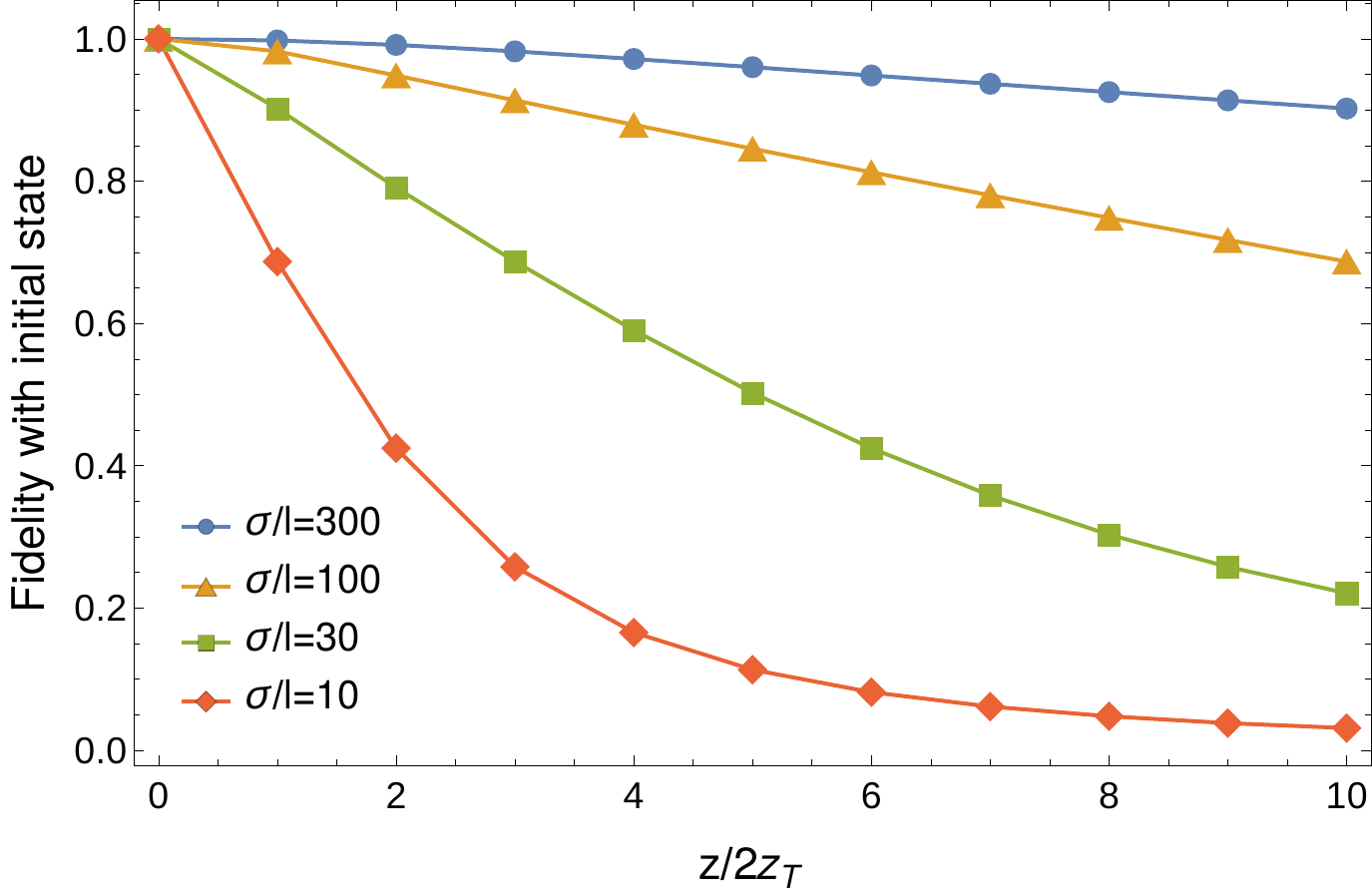}
\caption{ Fidelity of the initial state $|0_D\rangle$ with respect to itself after evolution by a distance  $m\, 2z_T$ with $m$ an integer,  considering the deviations from the paraxial approximation and finite number o illuminated slits. The results where calculated numerically considering the propagation of a Gaussian beam of width $\sigma$ at the grating and effectivelly illuminating  $l$ slits. While this fidelity decays rapidly for small values of $\sigma$ it is well preserved for values of the order of $10^2$ illuminated slits, as it is the case for reported experiments like the one in \cite{case09}. \label{fig:fiderrors} }
\end{figure}

\section{Resistance to Errors}
\label{sec:error}

In this proposal we present a new way of quantum information processing using the propagation of single-particle wave functions, namely the transversal profile of single photons.  A set of orthogonal periodic wave functions are used to encode quantum information on a finite $D$-dimensional system.  Although developed in a different context, our encoding scheme is reminiscient of one of the first ideas for quantum information processing with light presented in the GKP proposal~\cite{gkp01}. There, periodic basis states were encoded in the quadrature variable wave functions of single-mode fields, and quantum computation is realized in this setting using linear optics, squeezing, and homodyne detection.    As mentioned by GKP,  difficult non-linear optical processes are necessary in order to produce the initial non-Gaussian states, though there has been at least one proposal to approximately produce the GKP basis states \cite{vasconcelos10}. GKP demonstrated the advantages of periodic encoding in finite dimensions in terms of the feasibility of implementing quantum error correction codes in this discrete system.   With proper encoding, the GKP scheme is robust to displacements in phase space.  This is a characteristic that our realization directly inherits in principle, and straightforward adaptation of GKP's scheme for fault-tolerant codes should be possible.

Our proposal represents a feasible physical system with feasible tools for quDit manipulation exploiting the Talbot effect. Nevertheless, the perfect working mechanism of the Talbot effect relies on two important approximations: paraxiality, and infiniteness of the grating to generate the perfectly periodic combs.  In this respect Berry and Klein \cite{berry96}, have shown that departing from these approximations translates into a blurring of the Talbot images, which in our terms would mean a loss of fidelity of states and gates. We study the influence of these two approximations numerically by considering a Gaussian beam whose width covers $N$ slits, by propagating an initial state for several Talbot distances thus revealing  non-paraxial effects. We measure the fidelity of an initial state with respect to its own revival after a propagation $m$ times $2z_T$. The results are shown in Fig. \ref{fig:fiderrors}. For currently available gratings and laser beams, we estimate that for about 10 Talbot lengths the fidelity is higher than $0.9$ (blue curve) and thus the effects of paraxial approximation and infinite number of slits are well justified in this range.

\section{conclusions}
\label{sec:conc}

In this paper we have developed a new proposal for quantum information processing by means of the Talbot effect with individual photons and the help of some common linear optical elements. The Talbot gates have a rich mathematical structure, that of the circulant matrices, and are closely related to the discrete Fourier transform.  Interpretation of the Talbot effect as a quantum logic operation provides a beautiful if not interesting way to visualize computation through wave propatation and interference. Diagonal gates implemented by some phase structuring optical element such as commercial spatial light modulators, complete the tools for manipulation of single quDit gates.  In order to have the ability to implement entangling gates, a new spatially dependent beam splitter is proposed.  Construction of this device should be possible with existing technology.  Our proposal allows for the encoding and manipulation of multi-dimensional quantum systems, which are interesting for quantum information processing.

This framework seems to be a natural scenario for the quantum Fourier transform. All these elements put together render our quantum computation model universal for quantum systems with prime dimensions. For high dimensional systems, quantum information encoded in Talbot quDits would be robust to errors, since error correcting codes could be implemented according to the GKP proposal \cite{gkp01}. We believe that the ideas  we have developed here can also be relevant and useful for quantum information processing with other physical systems such as atomic beams \cite{chapman95} and Bose Einstein condensates \cite{nowak97}, where the Talbot effect has been observed.


\end{document}